%% file: main.tex
\documentclass[10pt,journal,twocolumn,twoside,letterpaper]{IEEEtran}

\makeatletter
\def\ps@headings{%
  \def\@oddhead{\scriptsize\sffamily
    \parbox[b]{\dimexpr\textwidth-2em\relax}{\raggedright\rightmark}\hfil\thepage}%
  \def\@evenhead{\scriptsize\sffamily\thepage\hfil\leftmark\hbox{}}%
  \def\@oddfoot{}\def\@evenfoot{}}
\let\ps@titlepagestyle\ps@headings
\makeatother
\usepackage[T1]{fontenc}
\usepackage{microtype}
\usepackage[hidelinks]{hyperref}
\usepackage{cite}
\usepackage{amsmath,amssymb}
\usepackage{graphicx}
\usepackage{booktabs}
\usepackage{siunitx}
\usepackage{placeins}

\newtheorem{lemma}{Lemma}
\newtheorem{proposition}{Proposition}

\DeclareMathOperator*{\argmax}{argmax}
\DeclareMathOperator*{\argmin}{argmin}

\title{Approximate Dynamic Programming for Link Reconfiguration to Reduce Malicious Input Controllability in Consensus Networks}

\author{
Nathaniel~Mailhot\textsuperscript{*}, Mohammad~Pirani, and Davide~Spinello
\thanks{N. Mailhot is with the Department of Mechanical and Aerospace Engineering, Carleton University, Ottawa, Canada. He was with the Department of Mechanical Engineering, University of Ottawa, Ottawa, Canada.}
\thanks{\textsuperscript{*}Corresponding author: \mbox{nathanielmailhot@cunet.carleton.ca}.}
\thanks{M. Pirani and D. Spinello are with the Department of Mechanical Engineering, University of Ottawa, Ottawa, Canada. Emails: \{mpirani, davide.spinello\}@uottawa.ca.}
}

\hypersetup{pdftitle={Approximate Dynamic Programming for Link Reconfiguration to Reduce Malicious Input Controllability in Consensus Networks},pdfauthor={Nathaniel Mailhot, Mohammad Pirani, Bao Nguyen, Davide Spinello}}

\begin{document}
\maketitle

\begin{abstract}
Communication links influence how strongly a malicious input at one sensor affects a consensus network.
We study how a defender can reduce this influence by selecting one additional link after observing the currently attacked sensor.
The attacked sensor changes over time, and vulnerability is measured by the trace of a finite-horizon controllability Gramian.
A prescribed local epsilon-greedy attacker response rule defines a discounted Markov decision process for the defender.
Exact dynamic programming (DP) provides an optimal policy using the full Gramian payoff matrix and attacker transition model.
We then introduce approximate dynamic programming (ADP) that learns from sampled attacker transitions and empirical Gramian costs estimated from measured impulse responses.
Under conditionally unbiased costs and standard sampling and learning-rate conditions, the learned action-value table converges to the exact DP solution.
The analysis uses Bellman contraction and bounds the effect of persistent Gramian estimation bias, establishing when measured responses recover optimal defender decisions without constructing the full payoff matrix.
In simulations on \num{20}- and \num{100}-node graphs, empirical ADP rollout costs are within \qty{0.23}{\percent} and \qty{1.76}{\percent}, respectively, of exact DP.
In a two-dimensional intruder tracking example, both policies reduce network-wide root-mean-square tracking error spread by approximately \qty{25}{\percent} relative to an unmitigated attack.
\end{abstract}


\section{Introduction}

\IEEEPARstart{D}{etection} networks in military and civilian applications are generally dedicated to the estimation of the state of the trajectories of moving intruders. They are composed of topologically separated sensors that observe the target and exchange information over an underlying communication graph to maintain a consistent trajectory estimate \cite{liSummaryTargetDetection2021,robinMultirobotTargetDetection2016}.
This interaction can be modelled as a consensus process on a network, where sensors are nodes and communication links are edges \cite{olfati2007consensus}.
Should an attacker hijack a sensor node and inject a malicious signal into its local estimate, the compromised node can pull the network away from agreement and degrade overall tracking estimate quality across all nodes \cite{pirani2019attackresilient}.
When the network's defender observes the attack, which new communication links between nodes should be activated to best reduce the network's vulnerability to the attack?

We consider the simple case where the defender only activates a single link in response to a single attacked node at discrete time events with constant intervals.
At a given decision stage, the set of network communication links consists of the fixed set of edges of the base graph, with the addition of one candidate communication link activated in response to the previous attack. In response to the new attack, a new candidate communication link is selected to replace the previous one, so that the total number of edges is constant through the process, although the topology is time-varying.
The attacker follows a local $\epsilon$-greedy response policy that moves toward higher vulnerability nearby nodes as a result of the topology changes caused by the defender's actions.
Hence, with one node compromised, the defender must choose which additional edge should be active at the next stage to reduce the network's vulnerability to the injected signal.
For a single decision stage, this interaction can be represented as a zero-sum Stackelberg game: the attacker selects a target node, and the defender observes that choice before selecting an additional communication link.

Controllability Gramians have previously been applied to various forms of network design. Prior approaches include using Gramian measures to rank static edge additions and weight adjustments to enhance controllability \cite{chanekar2021energytransfer}, and formulating static Stackelberg games where a defender implements self-feedback loops prior to an attacker selecting compromised nodes \cite{pirani2019consensusgame}. Additional methods involve estimating topology gradients from offline trajectory data for static controllability maximization, or employing online reinforcement learning to tune compensation gains under changing communication topologies \cite{banno2023offlinegradient,lu2026onlinecontrollability}.


In contrast, the present work considers a scenario where the compromised node evolves over successive decision stages, and the defender selects an added communication link after observing the current attack. Because this chosen link also alters the transition probabilities of the next attacked node, the framework gives rise to the sequential defender problem investigated under a specified attacker response law.

The optimal defender policy can be solved exactly via dynamic programming (DP) after computing the finite-horizon controllability Gramian cost for every attacked node and added edge pair.
This defines the controllability Gramian payoff matrix used by the defender.
The exact workflow then solves the Bellman equation over the attacked-node state, with the added link as the defender action.
The full payoff matrix must be available before the DP step, although its construction can exploit the symmetry of the defended Laplacians.
This motivates a complementary approximate dynamic programming (ADP) formulation in which the defender learns from sampled transitions and response-based costs.
ADP methods address infeasible exact dynamic programming problems by replacing exhaustive value-function computation with value approximation and sampled updates \cite{Lewis2012,SuttonBarto2018,Kiumarsi2014,AlTamimi2007}.
Here, ADP is used to learn an approximate action-value function for the defender from sampled interactions with the attacked network.
Each update uses one selected defender edge, one sampled next attacked node, and one empirical controllability Gramian cost estimated from the measured network response in place of a stored payoff matrix entry \cite{Lall1999Empirical}.
We establish conditions under which the sampled ADP updates recover the action-value function of exact DP and hence optimal defender decisions.
The analysis identifies the Bellman operator associated with the conditional mean of the empirical update, establishes its contraction in the infinity norm, and bounds the effect of persistent cost estimation error.
The simulations evaluate this recovery through policy agreement, action-value regret, and discounted rollout cost, together with computation time and the defensive effect in a two-dimensional airspace intruder tracking example.

The broader problem is sequential network reconfiguration under uncertainty, formulated here as a finite discounted MDP.
The procedure is to define the observed state and feasible changes, quantify disturbance effects, and evaluate each choice through its immediate and future costs.
With a network model and transition law, the resulting problem is solved by exact DP.
With measured impulse responses and observed transitions, ADP can recover an optimal policy without an explicit dynamics model for its updates, under the stated sampling and estimation conditions.

The main contributions of this paper are summarized as follows:
\begin{itemize}
    \item {\bf Sequential Network Reconfiguration:} We formulate link defence against a moving attacker as an instance of sequential network reconfiguration under uncertainty. A finite discounted Markov decision process combines the available link changes, their vulnerability costs, and the prescribed local $\epsilon$-greedy attacker response.
    \item {\bf Exact Controllability Benchmark:} We establish an exact optimal defender policy using Dynamic Programming based on the exhaustive offline construction of a finite-horizon controllability Gramian payoff matrix.
    \item { \bf Empirical ADP and Decision Recovery:} We learn link defence from measured impulse responses without constructing the full payoff matrix. We establish sufficient conditions for recovering optimal defender decisions with conditionally unbiased costs and with sufficiently small persistent Gramian estimation bias.
\end{itemize}

The remainder of the paper is organized as follows: Section~\ref{sec:problem} formulates the consensus network defence problem, Section~\ref{sec:exact} presents the exact DP solution, Section~\ref{sec:adp} develops the online controllability Gramian estimate and ADP solution, Section~\ref{sec:results} reports simulation results for an air-defence sensor network, and Section~\ref{sec:conclusion} summarizes our findings and provides future directions for the work.

\section{Problem Formulation}
\label{sec:problem}

We formulate the defender's link activation problem for a consensus network in which a moving attacker compromises one node at a time.
The formulation specifies a nominal consensus network, the defender's admissible link activations, how the compromised sensor changes over time, and the controllability payoff used to evaluate each attack and defence action pair.
The resulting interaction defines a one-stage attacker-defender game, and the specified attacker transition law leads to the sequential defender problem solved in the following sections.

\subsection{Network Dynamics and Attack}

Let
\begin{equation}
\mathcal{G}_0=(\mathcal{V},\mathcal{E}_0),
\qquad
\mathcal V=\{1,\ldots,n\},
\end{equation}
be an undirected communication graph that we call the base graph, with adjacency matrix $A_0$ and Laplacian
\begin{equation}
L_0=\Delta_0-A_0,
\end{equation}
where $\Delta_0$ is the degree matrix of $A_0$. The topology of this graph is time-invariant. 
Each sensor carries a scalar consensus state $x_i(t)$, and the stacked network state is $x(t) = \left(x_1,\ldots,x_n \right) \in\mathbb R^n$.
In the absence of attack or reconfiguration, the sensor network evolves as
\begin{equation}
\dot x(t)=-L_0x(t).
\label{eq:consensus}
\end{equation}

The attacker compromises one sensor node at a time.
Let
\begin{equation}
f_k\in\mathcal{F},\qquad \mathcal{F} \equiv \mathcal{V},
\end{equation}
denote the attacked node at time stage $k$.
The malicious input enters through the compromised sensor, so the attacked network dynamics are
\begin{equation}
\dot x(t)=-L_0x(t)+e_{f_k}u_a(t),
\label{eq:attacked_base_dynamics}
\end{equation}
where $e_{f_k}$ is the $f_k$th canonical basis vector in $\mathbb{R}^n$ and $u_a(t)$ is the scalar malicious input.

The greedy component of its motion may remain at the current node or move to a nearby node connected through $\mathcal{G}_0$.
Let $d_{\mathcal{G}_0}(i,j)$ denote the graph distance on the base graph, that is the minimum number of edges between nodes $i$ and $j$ \cite{godsil2001algebraic}, and let $r_A$ be the attacker mobility radius.
The admissible next attacked node set is
\begin{equation}
\mathcal{M}(f_k)
:=
\{j\in\mathcal{V}:\ d_{\mathcal{G}_0}(f_k,j)\le r_A\},
\label{eq:attacker_mobility_set}
\end{equation}
which includes $f_k$ itself.
For the greedy choice of the response law, the attacker transition satisfies
\begin{equation}
f_{k+1}\in\mathcal{M}(f_k).
\label{eq:attacker_local_motion}
\end{equation}
The exploratory branch of \eqref{eq:pff} also permits transitions outside this local set.

\subsection{Defender Action and Controllability Gramian Payoff}

At each decision stage, the defender activates one communication link that does not belong to the base graph.
The admissible defender action set is
\begin{equation}
\mathcal K
:=
\bigl\{\{i,j\}:i,j\in\mathcal{V},\ i<j,\ \{i,j\}\notin\mathcal{E}_0\bigr\},
\end{equation}
with $m=|\mathcal K|$ candidate links.
For an action $a=\{i,j\}\in\mathcal K$, the defended Laplacian is
\begin{equation}
L_a=L_0+(e_i-e_j)(e_i-e_j)^\top.
\label{eq:defended_lap}
\end{equation}
Here $e_i$ and $e_j$ are the $i$th and $j$th canonical basis vectors in $\mathbb R^n$.
Only one added link is active at any decision stage, so the defender's action fully determines the network topology for that stage.

For attacked node $f$, defender action $a$, and the time $T_{\mathrm{gram}}>0$, define the finite-horizon controllability Gramian \cite{wilsonj.rughLinearSystemTheory1996}
\begin{equation}
W_c(L_a,e_f)
=
\int_0^{T_{\mathrm{gram}}}
e^{-L_a t} e_f e_f^\top e^{-L_a^\top t}\,\mathrm{d}t.
\label{eq:gramian_def}
\end{equation}
Here $W_c(L_a,e_f)\in\mathbb R^{n\times n}$ and $e_f$ is the $f$th canonical basis vector.
The related scalar vulnerability cost is
\begin{equation}
J_c(f,a)=\operatorname{tr}W_c(L_a,e_f).
\label{eq:payoff}
\end{equation}
A larger value of $J_c(f,a)$ indicates that the malicious input at node $f$ can produce a larger aggregate state response over the controllability Gramian horizon under defended topology $L_a$.

\subsection{Single-Stage Stackelberg Game}

For a single decision stage, the interaction is a finite zero-sum Stackelberg game.
The attacker first selects the compromised node $f\in\mathcal F$, and the defender observes this choice before selecting an active edge $a\in\mathcal K$.
The payoff is the controllability vulnerability $J_c(f,a)$.
The attacker seeks to increase this payoff, while the defender seeks to reduce it.
Therefore
\begin{equation}
\max_{f\in\mathcal{F}}\min_{a \in \mathcal K }J_c(f,a).
\label{eq:stage_game}
\end{equation}
The outer maximization represents the attacker's choice of node, while the inner minimization represents the defender's choice of link after observing that node.

For a fixed attacked node, the defender's best response is
\begin{equation}
a^\star(f) \in \argmin_{a\in\mathcal K}J_c(f,a).
\label{eq:best_response}
\end{equation}

\subsection{Sequential Defender Decision}

Equation~\eqref{eq:stage_game} describes the single-stage game.
For the sequential problem, we prescribe the attacker's local $\epsilon$-greedy response rule and optimize only the defender's policy.
At stage $k$, the defender observes the compromised node $f_k$ and selects a link $a_k$.
The attacker then moves to $f_{k+1}$ according to the prescribed rule under the selected link.
Thus, the attacked node changes over time, but the response rule remains fixed.
The resulting defender problem is a discounted Markov decision process (MDP) with action-dependent transition probabilities.

At decision stage $k$, the MDP state is
\begin{equation}
s_k=f_k\in\mathcal F,
\label{eq:state}
\end{equation}
and the defender selects $a_k\in\mathcal K$.
The attacker then moves to $f_{k+1}$ under the selected link.
The resulting cost is
\begin{equation}
g(f_{k+1},a_k)=J_c(f_{k+1},a_k).
\label{eq:stage_cost}
\end{equation}

For a selected defender action $a$, the attacker assigns to each admissible next node $j\in\mathcal{M}(f)$ the vulnerability score $J_c(j,a)$.
The greedy attacker response set is
\begin{equation}
\mathcal{R}_A(f,a)
=
\argmax_{j\in\mathcal{M}(f)}J_c(j,a).
\label{eq:attacker_greedy_set}
\end{equation}
For the greedy choice, ties are resolved by choosing the smallest node index.
With exploration parameter $\epsilon_A\in[0,1]$, the attacked node transition probability is
\begin{equation}
p_a(f,f')
=
 \left\{ 
 \begin{array}{ll}
1-\epsilon_A+\frac{\epsilon_A}{n},
& f'=\min\mathcal R_A(f,a),\\
\frac{\epsilon_A}{n},
& f'\ne\min\mathcal R_A(f,a).
\end{array}
\right.
\label{eq:pff}
\end{equation}
Thus, the attacker greedily moves to a high vulnerability admissible node with probability $1-\epsilon_A$ and makes a global exploratory move with probability $\epsilon_A$.

For a defender policy $\pi:\mathcal F\rightarrow\mathcal K$, with $a_k=\pi(f_k)$, the discounted cost is
\begin{equation}
V^\pi(f)
=
\mathbb E^\pi
\left[
\sum_{k=0}^{\infty}\gamma^k g(f_{k+1},a_k)
\mid f_0=f
\right],
\label{eq:discounted_objective}
\end{equation}
where $\mathbb E^\pi[\cdot]$ denotes expectation under policy $\pi$ and transition law \eqref{eq:pff}, and $\gamma\in(0,1)$ is the discount factor.
The defender seeks a policy that minimizes \eqref{eq:discounted_objective} under the transition law \eqref{eq:pff}.

\section{Dynamic Programming Solution}
\label{sec:exact}

The exact solution computes the optimal defender policy for the Markov decision process defined in Section~\ref{sec:problem}.
This requires two offline computations.
The first computation evaluates the controllability payoff for every attacked node and defender action pair, forming the payoff matrix
\begin{equation}
\mathcal{J}
=
\bigl[J_c(f,a)\bigr]_{f\in\mathcal{F},\ a\in\mathcal K}
\in\mathbb R^{n\times m}.
\label{eq:payoff_matrix}
\end{equation}
The second computation solves Bellman's equation induced by the discounted objective in \eqref{eq:discounted_objective}.

\subsection{Payoff Matrix Construction}

For each attacked node $f\in\mathcal{F}$ and defender action $a\in\mathcal K$, the finite-horizon controllability Gramian can be obtained from the local form of \eqref{eq:gramian_def}
\begin{equation}
\dot W(t)=-L_aW(t)-W(t)L_a^\top+e_fe_f^\top,
\qquad
W(0)=0,
\label{eq:gramian_ode}
\end{equation}
where $W(t)\in\mathbb R^{n\times n}$ is the controllability Gramian state.
The payoff is
\begin{equation}
J_c(f,a)=\operatorname{tr}W(T_{\mathrm{gram}}).
\label{eq:payoff_from_ode}
\end{equation}
Here, $T_{\mathrm{gram}}>0$ is the time horizon over which the controllability Gramian is evaluated.

The payoff matrix in \eqref{eq:payoff_matrix} contains $nm$ entries, but these entries need not be computed independently.
Writing $T=T_{\mathrm{gram}}$, cyclic invariance of the trace gives
\begin{equation}
J_c(f,a)=e_f^\top\!\left(\int_0^T e^{-L_a^\top t}e^{-L_at}\,dt\right)e_f.
\label{eq:shared_trace_cost}
\end{equation}
Thus, one shared matrix computation for each defender action supplies all attacked-node costs.
For the symmetric Laplacians considered here, let $L_a=U_a\Lambda_aU_a^\top$, with eigenvalues $\lambda_{a,i}\ge0$ and orthonormal eigenvectors $u_{a,i}$.
Then
\begin{equation}
\begin{aligned}
J_c(f,a)&=\sum_{i=1}^n (u_{a,i})_f^2 h_T(\lambda_{a,i}),\\
h_T(\lambda)&=
\begin{cases}
\dfrac{1-e^{-2\lambda T}}{2\lambda},&\lambda>0,\\[4pt]
T,&\lambda=0.
\end{cases}
\end{aligned}
\label{eq:spectral_trace_cost}
\end{equation}
One dense eigendecomposition costs $O(n^3)$, and evaluating all $n$ costs from its eigenvectors costs $O(n^2)$, giving a shared payoff construction cost of $O(mn^3)$.
The zero-eigenvalue limit makes this formula well defined for consensus Laplacians without taking a matrix inverse.
The separate-pair implementation retained in the original timing comparison instead repeats a Gramian calculation for each $(f,a)$ pair.
Its $O(mn^4)$ dense-work proxy describes that implementation, rather than a necessary cost of exact DP. It also suppresses numerical integration work.

\subsection{Bellman's Equation}

Let $V^\star(f)=\min_\pi V^\pi(f)$ denote the optimal discounted value from attacked node $f$.
With the transition cost specified in Section~\ref{sec:problem}, the standard discounted Bellman equation \cite{Bertsekas2012} is
\begin{equation}
V^\star(f)
=\min_{a\in\mathcal K}\sum_{f'\in\mathcal F}
p_a(f,f')\left[J_c(f',a)+\gamma V^\star(f')\right].
\label{eq:bellman_exact}
\end{equation}
Here $f$ is the current attacked node, $a$ is the selected defender link, and $f'$ is the next attacked node.
The cost $J_c(f',a)$ is incurred under the selected link, followed by the discounted continuation value at $f'$.

The exact defender policy is
\begin{equation}
\pi^\star(f)
\in\argmin_{a\in\mathcal K}\sum_{f'\in\mathcal F}
p_a(f,f')\left[J_c(f',a)+\gamma V^\star(f')\right].
\label{eq:exact_policy}
\end{equation}
One exact DP iteration applies \eqref{eq:bellman_exact} over all attacked-node states $f\in\mathcal F$, evaluating every candidate action $a\in\mathcal K$.
After the admissible attacker motion sets $\mathcal{M}(f)$ and greedy attacker responses have been formed from the base graph and payoff matrix, each Bellman iteration has cost $O(nm)$.
Let $I_{\mathrm{DP}}$ denote the number of value-iteration steps.
The total computational cost of these steps is
\begin{equation}
O(I_{\mathrm{DP}}nm).
\label{eq:dp_compute_count}
\end{equation}
With the shared spectral construction, these two steps have total cost $O(mn^3+I_{\mathrm{DP}}nm)$, in addition to constructing the attacker motion sets and greedy responses.

\section{Approximate Dynamic Programming Solution}
\label{sec:adp}

The ADP update uses sampled empirical observations of the network's controllability to approximate the defender's optimal action-value function.
The update itself does not require the model-derived payoff matrix $\mathcal{J}$ in \eqref{eq:payoff_matrix}, provided that costs and attacker transitions can be observed or sampled.
Instead, it learns an approximate action-value function from observed attacker transitions and controllability Gramian cost estimates obtained from responses measured along the network's trajectories.
This provides a route to incremental learning from responses. A runtime advantage over an efficient exact solver is not implied.

\subsection{Sampled Action Value Update}

Let
\begin{equation}
\hat Q_k:\mathcal F\times\mathcal K\rightarrow \mathbb{R}
\end{equation}
denote the approximate action-value function after an update at stage $k$.
The entry $\hat{Q}_k(f,a)$ estimates the discounted cost of selecting defender action $a$ when the current attacked node is $f$.
The initial approximation is
\begin{equation}
\hat{Q}_0(f,a)=q_0,
\qquad f\in\mathcal{F},\ a\in\mathcal K,
\label{eq:adp_init}
\end{equation}
where $q_0$ is a constant.

The exact action-value function associated with \eqref{eq:bellman_exact} is
\begin{equation}
\begin{aligned}
Q^\star(f,a)
&=\sum_{f'\in\mathcal F}p_a(f,f')
\left[J_c(f',a)+\gamma V^\star(f')\right]\\
&=\sum_{f'\in\mathcal F}p_a(f,f')
\left[J_c(f',a)+\gamma\min_{b\in\mathcal K}Q^\star(f',b)\right].
\end{aligned}
\label{eq:qstar_from_v}
\end{equation}
The optimal defender policy in terms of the action-value function is obtained by
\begin{equation}
\pi^\star(f)
\in
\argmin_{a\in\mathcal K}Q^\star(f,a).
\label{eq:qstar_policy}
\end{equation}
The table $Q^\star$ is the target of the ADP approximation.
It is computed offline for evaluation of policy agreement and regret. The learner does not use its entries in the update.

At update $k$, the ADP method samples an attacked node $f_k$ and a defender action $a_k$.
With $f_k$ and $a_k$ fixed, the next attacked node is sampled according to
\begin{equation}
\Pr(f_{k+1}=f')=p_{a_k}(f_k,f'),
\qquad f'\in\mathcal F.
\label{eq:adp_sample_next}
\end{equation}
The sampled transition replaces the full expectation over $f'\in\mathcal F$ used in the exact Bellman update.

\subsection{Empirical Controllability Gramian Cost Estimate}

Let $C_S\in\mathbb R^{s\times n}$ be a sensor selection matrix that maps the full network state to the $s$ measured sensor states available to the empirical estimator.
For the response associated with attacked node $f_{k+1}$ and active defended edge $a_k$, define
\begin{subequations}
\label{eq:empirical_response}
\begin{align}
y_k(t)=C_S\delta x_k(t),
&\qquad
\dot{\delta x}_k(t)=-L_{a_k}\delta x_k(t),\\
&\qquad
\delta x_k(0)=e_{f_{k+1}}
\end{align}
\end{subequations}
where $\delta x_k(t)\in\mathbb{R}^n$ is the impulse response state and $y_k(t)\in\mathbb{R}^s$ is the measured response.
Following the empirical controllability Gramian construction of \cite{Lall1999Empirical}, the observed scalar cost is the sampled response energy
\begin{equation}
c_k
=
\frac{n}{s}
\sum_{\ell=0}^{q-1}
\|y_k(\ell\Delta t)\|_2^2\Delta t
\label{eq:adp_scalar_cost}
\end{equation}
where $\ell$ is the sample index, $\Delta t$ is the response sampling interval, and $q$ is the number of samples used in the empirical response window.
When $C_S=I_n$ and $q\Delta t=T_{\mathrm{gram}}$, \eqref{eq:adp_scalar_cost} is a quadrature approximation of $J_c(f_{k+1},a_k)$.
At a fixed sampling interval, this estimate can have a persistent discretization error, even when every sensor is measured.
When only a subset of sensors is measured, \eqref{eq:adp_scalar_cost} estimates the response energy projected onto the measured sensor coordinates.
In either case, the ADP update uses a single scalar cost observation for the sampled state-action pair, rather than first evaluating all defender actions at that state.

The sampled ADP update is
\begin{align}
\hat Q_{k+1}(f_k,a_k)
&=
(1-\alpha_k)\hat Q_k(f_k,a_k) \nonumber \\
&\quad
+
\alpha_k
\left[
c_k+\gamma\min_{b\in\mathcal K}\hat Q_k(f_{k+1},b)
\right],
\label{eq:adp_update_sampled_row}
\end{align}
where $\alpha_k\in(0,1]$ is the learning rate.
All entries not sampled at update $k$ remain unchanged
\begin{equation}
\hat Q_{k+1}(f,a)=\hat Q_k(f,a),
\qquad (f,a)\neq(f_k,a_k).
\label{eq:adp_unchanged_entries}
\end{equation}
Equation~\eqref{eq:adp_update_sampled_row} can be applied online because it requires only the selected defender action, the realized next attacked node, and the measured scalar response cost.
The ADP update uses one observation of the next attacked node, while exact DP averages over all possible next attacked nodes under the transition model.
\subsection{Recovery of the Exact DP Solution}
\label{sec:adp_recovery}

The following lemma specializes the standard discounted Bellman contraction result in Bertsekas \cite{Bertsekas2012} to the defender problem.
The following proposition connects this operator to the sampled empirical update using the standard convergence theory for tabular Q-learning \cite{Tsitsiklis1994Qlearning}.
Proofs are given in Appendix~\ref{app:recovery_proofs}.

\begin{lemma}[Bellman fixed point for the defender MDP]
\label{lem:bellman}
Fix $p_a(f,f')$ and let $\gamma\in(0,1)$.
Define Bellman's operator on $Q:\mathcal F\times\mathcal K\rightarrow\mathbb R$ by
\begin{equation}
(\mathcal TQ)(f,a)
=\sum_{f'\in\mathcal F}p_a(f,f')
\left[J_c(f',a)+\gamma\min_{b\in\mathcal K}Q(f',b)\right].
\label{eq:adp_bellman_operator}
\end{equation}
Then $\mathcal T$ is a $\gamma$-contraction in the infinity norm and has the unique fixed point $Q^\star=\mathcal TQ^\star$.
In particular, for any approximate table $\hat Q$,
\begin{equation}
\|\mathcal T\hat Q-Q^\star\|_\infty
\le\gamma\|\hat Q-Q^\star\|_\infty.
\label{eq:adp_exact_contraction}
\end{equation}
The optimal value and policy satisfy
\begin{equation}
V^\star(f)=\min_{a\in\mathcal K}Q^\star(f,a),
\label{eq:vstar_from_qstar}
\end{equation}
\begin{equation}
\pi^\star(f)\in\argmin_{a\in\mathcal K}Q^\star(f,a).
\label{eq:qstar_policy_proposition}
\end{equation}
If $\|\hat Q-Q^\star\|_\infty\le\eta$, any greedy policy $\hat\pi$ for $\hat Q$ has one-step action-value regret
\begin{equation}
Q^\star(f,\hat\pi(f))-\min_{a\in\mathcal K}Q^\star(f,a)
\le2\eta,\qquad f\in\mathcal F.
\label{eq:adp_regret_bound}
\end{equation}
\end{lemma}

For the following proposition, let $\mathcal H_k$ contain the learning history, $\hat Q_k$, $(f_k,a_k)$, and $\alpha_k$ before the next node and cost are sampled.
Given $\mathcal H_k$, the next node probabilities are $p_{a_k}(f_k,f')$.
Assume the costs have uniformly bounded conditional second moments and fixed conditional mean
\begin{equation}
\mathbb E[c_k\mid\mathcal H_k,f_{k+1}=f']
=\bar J_c(f',a_k).
\label{eq:adp_cost_mean}
\end{equation}
Almost surely, every state-action pair is updated infinitely often and its learning rates satisfy
\begin{equation}
\sum_{k:(f_k,a_k)=(f,a)}\alpha_k=\infty,
\qquad
\sum_{k:(f_k,a_k)=(f,a)}\alpha_k^2\le C_\alpha.
\label{eq:adp_stepsizes}
\end{equation}
Here $C_\alpha<\infty$ is a deterministic constant common to all pairs.

\begin{proposition}[Recovery of exact DP by empirical ADP]
\label{prop:adp_recovery}
Under these sampling conditions, let $\bar{\mathcal T}$ be the Bellman operator in \eqref{eq:adp_bellman_operator} with $J_c$ replaced by $\bar J_c$.
Then $\bar{\mathcal T}$ is a $\gamma$-contraction in the infinity norm with a unique fixed point $\bar Q^\star$.
The ADP update \eqref{eq:adp_update_sampled_row} satisfies
\begin{equation}
\Vert \hat Q_k-\bar Q^\star \Vert_\infty\rightarrow 0
\quad\text{almost surely}.
\label{eq:adp_convergence}
\end{equation}
For $\delta=\max_{f',a}|\bar J_c(f',a)-J_c(f',a)|$,
\begin{equation}
\|\bar Q^\star-Q^\star\|_\infty
\le\frac{\delta}{1-\gamma}.
\label{eq:adp_bias_bound}
\end{equation}
For conditionally unbiased costs, $\bar J_c=J_c$ and $\bar Q^\star=Q^\star$.
Every greedy policy is then optimal for all sufficiently large $k$, almost surely.
\end{proposition}

Under Proposition~\ref{prop:adp_recovery}'s assumptions, bias bounded by $\delta$ gives limiting action-value regret at most $2\delta/(1-\gamma)$.
Greedy decisions eventually become optimal if this bound is below every positive gap $Q^\star(f,a)-V^\star(f)$.

The greedy defender policy derived from the learned action-value function is
\begin{equation}
\hat\pi_k(f)
\in
\argmin_{a\in\mathcal K}\hat Q_k(f,a).
\label{eq:adp_policy}
\end{equation}
The minimization in \eqref{eq:adp_policy} is over the full defender action set $\mathcal K$.

Including the empirical response calculation in \eqref{eq:adp_scalar_cost}, $N_{\mathrm{emp}}$ empirical response estimates with $q$ response samples and $U_{\mathrm{ADP}}$ scalar action-value updates have computational cost
\begin{equation}
O\!\left(N_{\mathrm{emp}}q(|\mathcal E_0|+n)+U_{\mathrm{ADP}}\right).
\label{eq:adp_compute_count}
\end{equation}
Here $|\mathcal E_0|$ is the number of edges in the base graph.
Thus, the ADP computation replaces explicit construction of the $nm$-entry controllability Gramian payoff matrix with empirical controllability Gramian estimates and scalar action-value updates.

\section{Experimental Results}
\label{sec:results}

The following numerical simulations use connected small-world type base graphs generated by randomly rewiring sparse ring lattices with initial ring degree $d_0$ and rewiring probability $p_{\mathrm{rew}}$ \cite{watts1998smallworld}.
The first, small graph case uses $(n,d_0,p_{\mathrm{rew}})=(20,2,0.30)$ and $T_{\mathrm{gram}}=\num{10}$.
The second, large graph case uses $(n,d_0,p_{\mathrm{rew}})=(100,4,0.25)$ and $T_{\mathrm{gram}}=\num{8}$.
Appendix~\ref{app:vulnerability_plot} illustrates the vulnerability costs for the \num{20}-node graph.

For the sequential attack model, the attacker uses the local graph set \eqref{eq:attacker_mobility_set} with radius $r_A=2$ measured on the base graph.
Thus, when taking the greedy part of its response, the attacker can remain at the current sensor or move to any sensor within two hops along the base graph.
The attacker uses \eqref{eq:pff} with $\epsilon_A=\num{0.10}$, so the greedy response is restricted to this local neighbourhood, while the exploration term assigns probability to every node in the network.
The discount factor is $\gamma=\num{0.97}$.

The ADP implementation samples attacked-node/action pairs uniformly and uses a per pair learning rate $\alpha=\alpha_0j^{-0.62}$ on the $j$th visit, with $\alpha_0=0.35$ and $0.28$ for the \num{20}- and \num{100}-node cases, respectively.
Continued sampling with this schedule satisfies the visitation and learning rate conditions in Proposition~\ref{prop:adp_recovery}.
The reported runs use full sensor measurements and a fixed response integration step of \qty{0.04}{\second}. Their finite training and discretization errors are assessed against exact DP rather than assumed to vanish.

The numerical results test whether empirical ADP recovers the defender performance of exact DP under the same attacker model.
Table~\ref{tab:wall_clock} retains the original implementation timings on an Apple M4 Max MacBook Pro, together with their analytic work proxies.
The DP comparator forms payoffs separately for each attacked-node/action pair before value iteration. Its proxy is $nmn^3+I_{\mathrm{DP}}nm$.
The ADP proxy counts response calculations and scalar updates as $N_{\mathrm{emp}}q(|\mathcal E_0|+n)+U_{\mathrm{ADP}}$. These proxies are not measured operation counts or complete implementation costs.
In particular, the ADP proxy omits row-minimum rescans and diagnostics.
The shared construction in \eqref{eq:spectral_trace_cost} removes the repeated attacked-node computations from the exact comparator.
Consequently, the original timing ratio is specific to the implemented separate-pair baseline and does not establish an ADP speed advantage over structure-exploiting exact DP.

A subsequent saved run using the shared spectral construction gave exact DP totals of \qty{0.0523}{\second} and \qty{4.13}{\second}, compared with empirical ADP training times of \qty{2.81}{\second} and \qty{42.8}{\second}, for the \num{20}- and \num{100}-node cases, respectively.
The exact and learned policies in these saved runs agree with their original counterparts, but the runtime ordering is reversed.
The exact totals combine a median payoff-construction time with a separately timed DP solve. These records are implementation measurements, not end-to-end latency guarantees.

The empirical costs in these experiments are simulated impulse-response energies, and the learner caches them for reuse.
The reported runs acquire \num{3400} of \num{3400} and \num{474989} of \num{475000} distinct state-action costs, respectively, so they do not demonstrate sparse acquisition of the empirical cost table.
The simulator also constructs its greedy attacker responses from the exact payoff matrix before timed learning. This preparation is excluded from the reported ADP time.
The experiments therefore assess policy recovery with a fixed simulator, rather than an entirely payoff-free simulation pipeline or physical online acquisition time.

\begin{table}[!t]
\centering
\caption{Original implementation timings and analytic work proxies. The DP rows use separate-pair Gramian construction, not the shared spectral construction.}
\label{tab:wall_clock}
\small
\setlength{\tabcolsep}{4pt}
\begin{tabular}{@{}lrrr@{}}
\toprule
Method &
{Network} &
{\shortstack{Legacy work\\proxy}} &
{\shortstack{Wall-clock\\time (s)}}\\
\midrule
DP (separate pairs) & \num{20} nodes & \num{2.98e7} & \num{3.22}\\
DP (separate pairs) & \num{100} nodes & \num{4.75e11} & \num{6.88e3}\\
ADP & \num{20} nodes & \num{4.00e7} & \num{3.99}\\
ADP & \num{100} nodes & \num{2.85e10} & \num{58.6}\\
\bottomrule
\end{tabular}
\end{table}

The policy quality is summarized in Table~\ref{tab:policy_quality}.
The ADP policy is evaluated using the same attacker response law and rollout horizon as the exact DP policy.
The \num{20}-node and \num{100}-node ADP runs use \num{6.0e6} and \num{5.04e7} scalar ADP updates with cached empirical costs, respectively.
In these experiments, the empirical response costs closely matched the corresponding exact Gramian costs in the full-state measurement case.
Policy match with exact DP is counted when the selected ADP edge is within \qty{0.10}{\percent} one-step regret of the exact DP action-value optimum.
Under this definition, the final ADP policy match is \qty{100}{\percent} for the \num{20}-node case and \qty{82}{\percent} for the \num{100}-node case.
The remaining mismatches are concentrated in sparsely sampled state-action regions and in near-tie states, where several candidate links produce nearly identical action values and therefore negligible regret.
The final mean one-step action-value regret is below \qty{0.1}{\percent} in both graph sizes, and the ADP rollout cost is within \qty{0.23}{\percent} and \qty{1.76}{\percent} of exact DP.

\begin{table}[!t]
\centering
\caption{Discounted rollout costs.}
\label{tab:policy_quality}
\small
\setlength{\tabcolsep}{4pt}
\begin{tabular}{@{}l
S[table-format=3.0]
S[table-format=2.1]
S[table-format=2.1]
S[table-format=2.1]@{}}
\toprule
Case & {\shortstack{No added\\edge}} & {\shortstack{Random\\edge}} & {Exact DP} & {\shortstack{Empirical\\ADP}}\\
\midrule
\num{20} nodes & 54.4 & 50.9 & 33.7 & 33.8\\
\num{100} nodes & 12.9 & 12.8 & 10.6 & 10.8\\
\bottomrule
\end{tabular}
\end{table}

\begin{figure*}[!t]
\centering
\includegraphics[width=0.94\textwidth]{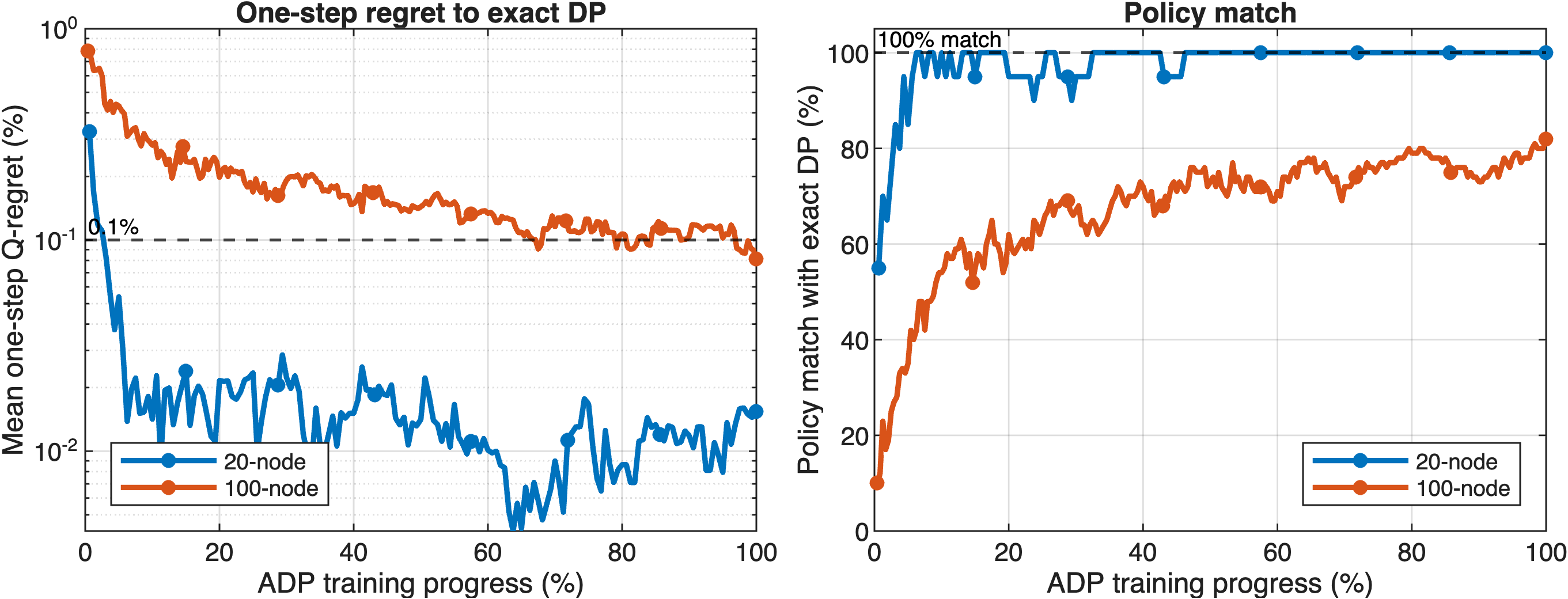}
\caption{Empirical ADP convergence diagnostics. Solid curves show medians over \num{10} independent ADP training runs, and shaded regions show the middle \qty{50}{\percent} of values across those runs. Left: mean one-step action value regret relative to the exact DP action-value function, with the dashed line marking \qty{0.1}{\percent}. Right: policy match with exact DP, where a match means the selected edge is within \qty{0.10}{\percent} one-step regret of the exact DP optimum. The dashed line marks \qty{100}{\percent} match.}
\label{fig:adp_regret}
\end{figure*}

Fig.~\ref{fig:adp_regret} shows how the empirical ADP policy changes as observations accumulate.
The regret plot shows decreasing one-step action-value regret relative to exact DP, while the policy match plot shows the percentage of attacked node states whose selected edge is within \qty{0.10}{\percent} one-step regret of the exact DP optimum.

\input{source_reports/airdefence_attack_sequence_table}

\subsection{Two-Dimensional Air-Defence Tracking}

The same \num{20}-node topology and frozen exact DP and empirical ADP policies were then evaluated in a two-dimensional airspace intruder tracking simulation using the attacked consensus network defined in Section~\ref{sec:problem}.
Let $E(t)\in\mathbb R^{n\times 2}$ stack the longitude and latitude tracking errors of the sensors.
For attacked node $f$ and active edge $a$, the error dynamics are
\begin{equation}
\dot E(t)=-L_aE(t)+e_fu_a(t)^\top+\Xi(t),
\label{eq:airdef_2d}
\end{equation}
where $u_a(t)\in\mathbb R^2$ is the injected attack signal and $\Xi(t)$ is process noise.
The attacked node path is generated once using the same $\epsilon$-greedy transition law \eqref{eq:pff}.
To compare the tracking cases under the same attack locations, we first sample one attacked-node sequence from \eqref{eq:pff}, using the exact DP policy only to define the nominal transition probabilities.
At decision stage $k$, the nominal defender action is $\bar a_k=\pi^\star(\bar f_k)$. With $\bar f_k$ and $\bar a_k$ fixed, the next node is drawn with probability $\Pr(\bar f_{k+1}=f')=p_{\bar a_k}(\bar f_k,f')$ for $f'\in\mathcal F$.
This fixed attacked node sequence is then used for all three tracking cases under attack.
In the attack with no defender case, the same attacked nodes are used, but no defensive link activation is applied.

Table~\ref{tab:airdefence_attack_sequence} lists this sequence.
The sequence uses $\epsilon_A=\num{0.10}$, decision interval \qty{1}{\second}, integration step $\Delta t=\qty{5.0e-2}{\second}$, and process noise level $\sigma_{\mathrm{process}}=\num{0.18}$ deg/$\sqrt{\mathrm{s}}$.
The four simulated cases are no attack, attack with no added edge, attack with the exact DP defender, and attack with the empirical ADP defender.

\begin{figure*}[!t]
\centering
\includegraphics[width=0.94\textwidth]{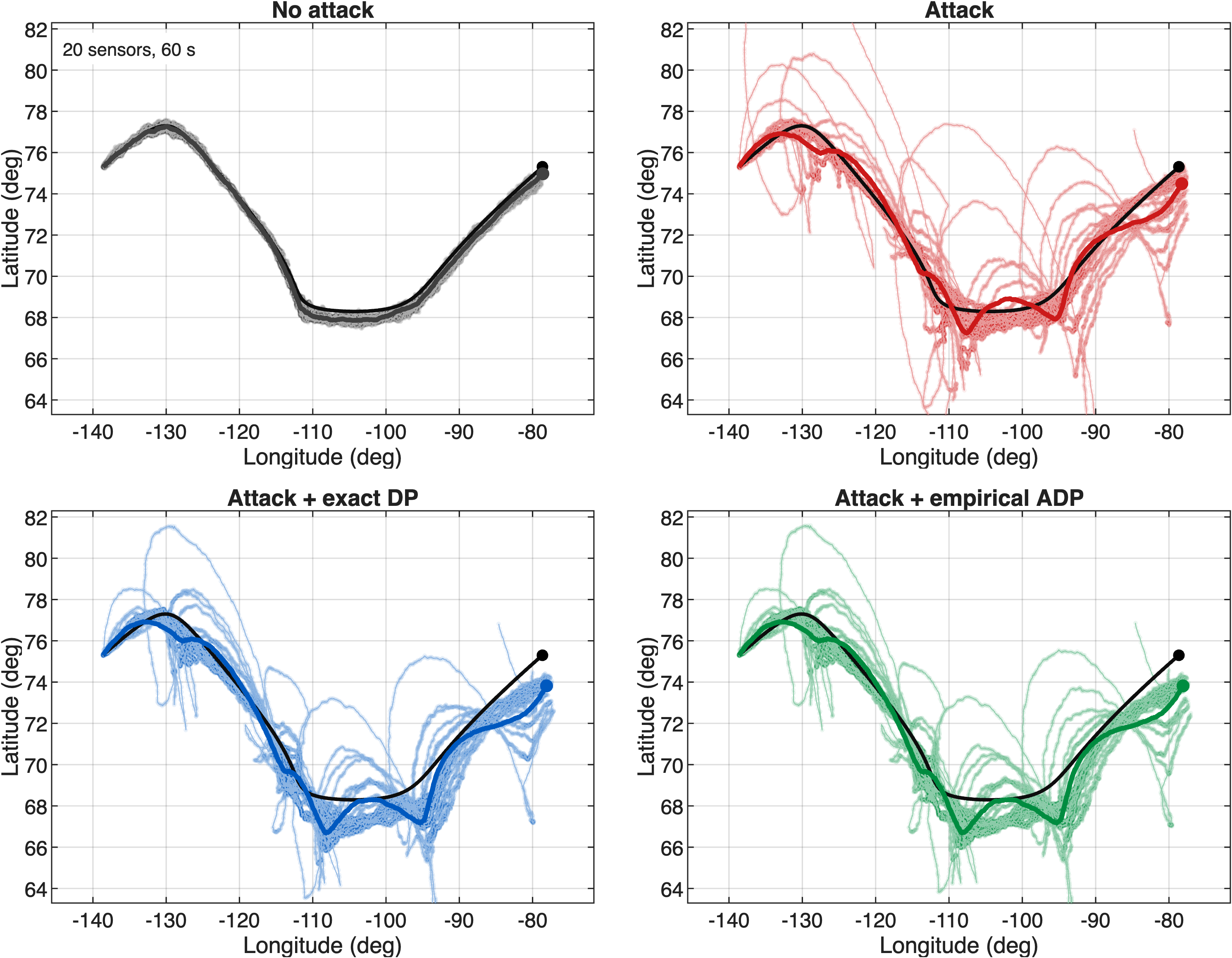}
\caption{Longitude-latitude trajectories of the \num{20} sensor node estimates over \qty{60}{\second}. The black curve is the true intruder path, the darker colored curves are the mean network estimate, and the faint paths show individual sensor estimates from all 20 nodes in each case.}
\label{fig:airdefence_trajectories}
\end{figure*}

Fig.~\ref{fig:airdefence_trajectories} shows the trajectories of all \num{20} sensor estimates in these four cases.
The no attack estimates remain tightly clustered around the common intruder track, while the attack only case produces visibly dispersed sensor paths.
Both defender policies reduce this dispersion and recover a concentrated tracking pattern.

\begin{figure}[!htbp]
\centering
\includegraphics[width=\linewidth]{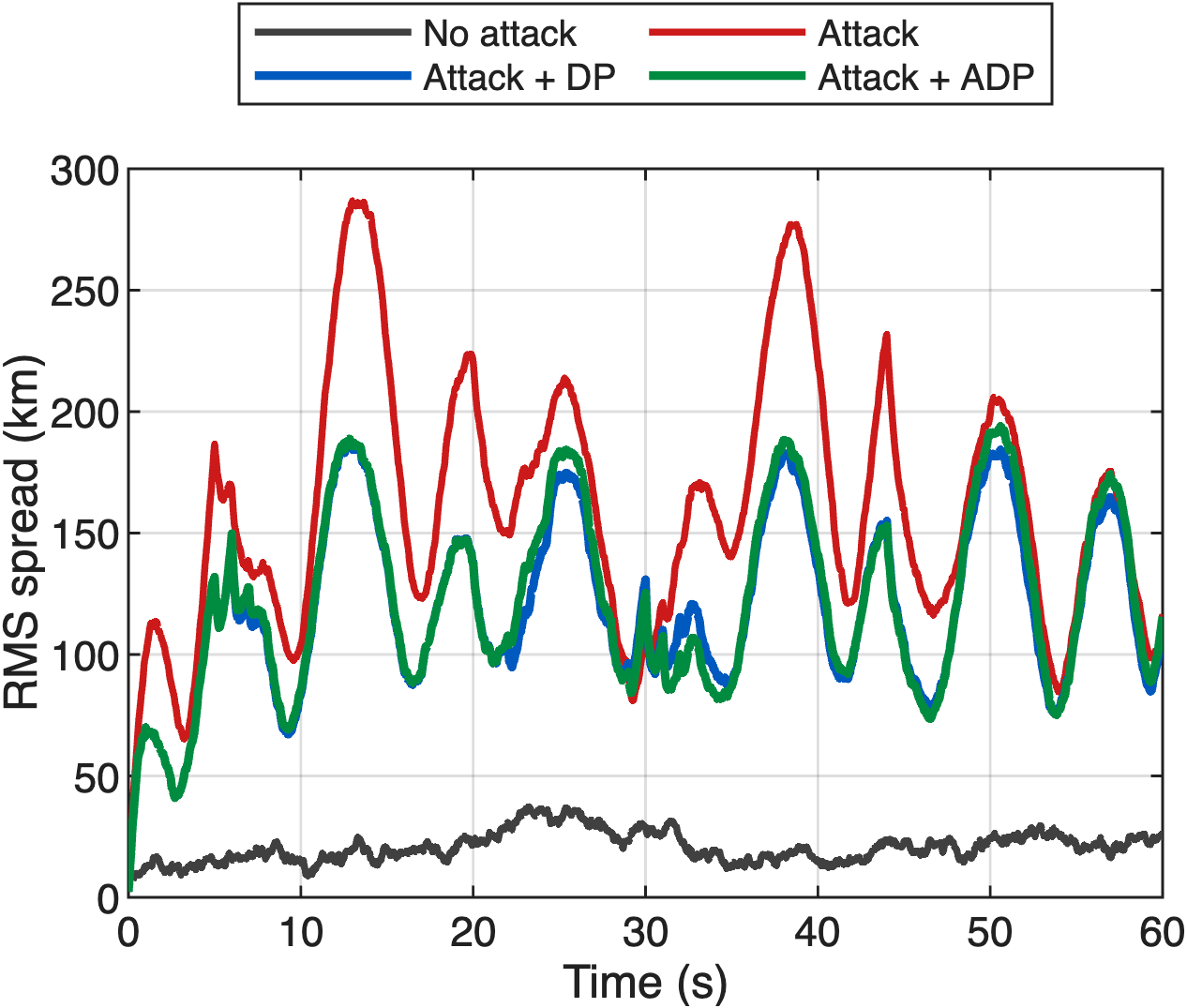}
\caption{Two-dimensional air-defence tracking spread under no attack, attack with no added edge, attack with exact DP defence, and attack with empirical ADP defence.}
\label{fig:airdefence_dp_adp}
\end{figure}

Fig.~\ref{fig:airdefence_dp_adp} reports the RMS spread of the sensor estimates over \qty{60}{\second}.
The time-averaged RMS spread is \qty{20.6}{\kilo\metre} without attack and increases to \qty{160}{\kilo\metre} under the unmitigated attack.
The exact DP defender reduces this value to \qty{120}{\kilo\metre}, a \qty{25.2}{\percent} reduction relative to the unmitigated attack.
The ADP defender gives a similar value of \qty{122}{\kilo\metre}, corresponding to a \qty{24.0}{\percent} reduction.

\section{Conclusion}
\label{sec:conclusion}

This paper studied defensive link reconfiguration for an air-defence sensor network that uses consensus coupling to maintain a coherent estimate of a moving intruder.
A cyber attacker was assumed to compromise one sensor at a time and inject a malicious input whose effect propagates through the network.
The defender observes the attacked sensor and activates one additional communication link to reduce the resulting vulnerability.
At a single decision stage, the interaction is a zero-sum Stackelberg game whose payoff is the trace of a finite-horizon controllability Gramian, which measures how strongly an injected signal can influence the network state.
Across repeated decision stages, the specified attacker response law induces the discounted defender MDP solved in this paper.

The analysis establishes when empirical ADP recovers the defender solution computed by exact DP.
The conditional mean of the sampled update is governed by a Bellman operator that is a contraction in the infinity norm.
With conditionally unbiased cost measurements, repeated sampling of every state-action pair, and the stated learning rate conditions, the learned table converges to $Q^\star$ and its greedy decisions become optimal.
For a persistent cost estimation error bounded by $\delta$, the limiting action-value error is at most $\delta/(1-\gamma)$.
These results connect the measured Gramian costs and sampled updates directly to the exact defender solution.

The exact DP solution provides a benchmark policy using the complete payoff matrix, which can be constructed efficiently by sharing computations across attacked nodes.
The empirical ADP update retains the same defender action space, attacker response law, and discounted objective, while using sampled response costs and action-value updates.
Its rollout cost is within \qty{0.23}{\percent} and \qty{1.76}{\percent} of the exact benchmark on the \num{20}- and \num{100}-node graphs, respectively.
These results support policy recovery in the simulated setting.
The original timing comparison uses a separate-pair exact implementation and does not establish a speed advantage over optimized exact DP.

The tracking simulation demonstrates the practical effect of the frozen exact DP and empirical ADP policies in the attacked consensus network.
The exact DP and empirical ADP defenders reduce the time-averaged RMS spread by \qty{25.2}{\percent} and \qty{24.0}{\percent}, respectively, relative to the unmitigated attack.
Although the RMS spread is not the controllability objective used during policy computation, the similar reductions show that the empirical ADP policy retains the practical defensive effect of the exact benchmark.

The same decision structure can be used for other finite sets of admissible linear-network configurations, including changes in link weights or groups of links.
Configurations specify the linear dynamics and attacked locations specify input channels, allowing the same Gramian cost to cover directed or weighted networks and multidimensional node dynamics.
Potential applications are choosing communication configurations for microgrids under cyber attacks~\cite{Danzi2018SoftwareDefined} and robot teams with compromised members~\cite{Lee2025ResilienceAware}.
Such extensions must supply suitable response measurements, satisfy the finite-MDP and learning assumptions, and respect the application's safety and stability constraints.

Future work will consider attackers that optimize their movement over time, leading to a zero-sum Markov game rather than a fixed attacker response model.
Minimax ADP methods could then learn defender policies against strategically optimizing attackers without constructing the complete payoff matrix.
Methods that use empirical responses provide one possible way to evaluate candidate configurations for nonlinear systems \cite{Lall1999Empirical}, while recent nonlinear pinning control results illustrate the additional modelling and stability conditions required for coupled nonlinear networks \cite{DellaRossa2023Nonlinear}.
Because the measured response can depend on the operating condition and attack magnitude, a nonlinear extension would require a richer learning state, new convergence analysis, and stability analysis for switching configurations.
A continuous-time differential game could further represent attack signals, link switching times, and reconfiguration costs directly.
Other directions include partial observability of the attacked node, multiple simultaneous attacks, weighted link reconfiguration, and larger geographically realistic air defence networks.

\newpage
\appendices
\section{Bellman and ADP Recovery Proofs}
\label{app:recovery_proofs}
\input{adp_recovery_proofs}

\section{Illustration of the Vulnerability Cost}
\label{app:vulnerability_plot}
\begin{figure}[!htbp]
\centering
\includegraphics[width=\linewidth]{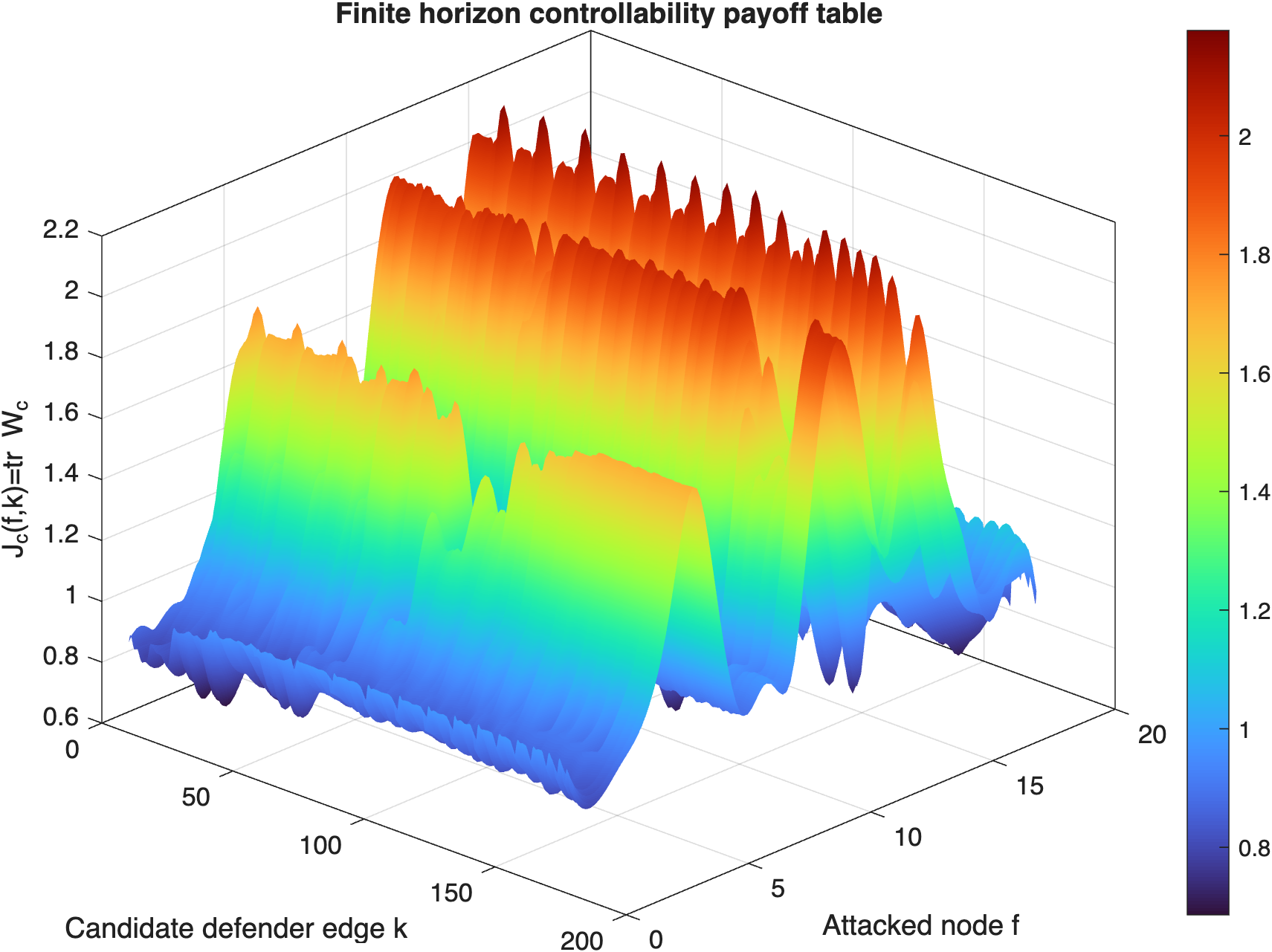}
\caption{Vulnerability cost $J_c(f,a)$ for the \num{20}-node graph. Lower values mean less vulnerability. The smooth surface is a spline interpolation between discrete node and link indices, used only for visualization.}
\label{fig:payoff_surface}
\end{figure}

Fig.~\ref{fig:payoff_surface} shows the discrete cost landscape underlying link selection.
At a fixed attacked node, the lowest-cost candidate link gives the defender's best single-stage response.
Taking these minima across nodes gives the best-response profile, whose largest value is the Stackelberg payoff in~\eqref{eq:stage_game}.
In this example, no single link is best for every attack location, making the need for location-dependent decisions visible.
DP uses this cost table and the attacker transition model in Bellman updates to compare link choices over time.
ADP learns the corresponding action values from sampled transitions and network responses.
\FloatBarrier

\section*{Acknowledgment}
The authors thank Dr. Bao Nguyen for helpful discussions and guidance in defining the problem and its application to airdefence tracking.

\bibliographystyle{IEEEtran}
\bibliography{bib}

\end{document}

%% file: source_reports/airdefence_attack_sequence_table.tex
\begin{table*}[!t]
\centering
\caption{Attacked-node sequence for all tracking cases.}
\label{tab:airdefence_attack_sequence}
\begin{minipage}[t]{.49\textwidth}
\centering
\scriptsize
\setlength{\tabcolsep}{3pt}
\begin{tabular}{@{}llllll@{}}
\toprule
Stage & Time (\si{\second}) & Atk node $f$ & Active $\kappa$ & Active edge & $\mathrm{tr}\,W_c$ \\ \midrule
\num{1} & $[\num{0},\num{1})$ & \num{6} & \num{19} & $(\num{2},\num{6})$ & \num{0.9603} \\
\num{2} & $[\num{1},\num{2})$ & \num{9} & \num{149} & $(\num{13},\num{15})$ & \num{0.9676} \\
\num{3} & $[\num{2},\num{3})$ & \num{13} & \num{149} & $(\num{13},\num{15})$ & \num{1.026} \\
\num{4} & $[\num{3},\num{4})$ & \num{16} & \num{150} & $(\num{13},\num{16})$ & \num{0.9604} \\
\num{5} & $[\num{4},\num{5})$ & \num{4} & \num{150} & $(\num{13},\num{16})$ & \num{0.9268} \\
\num{6} & $[\num{5},\num{6})$ & \num{4} & \num{84} & $(\num{6},\num{13})$ & \num{0.9648} \\
\num{7} & $[\num{6},\num{7})$ & \num{10} & \num{84} & $(\num{6},\num{13})$ & \num{0.9733} \\
\num{8} & $[\num{7},\num{8})$ & \num{11} & \num{149} & $(\num{13},\num{15})$ & \num{1.426} \\
\num{9} & $[\num{8},\num{9})$ & \num{18} & \num{150} & $(\num{13},\num{16})$ & \num{0.8456} \\
\num{10} & $[\num{9},\num{10})$ & \num{2} & \num{150} & $(\num{13},\num{16})$ & \num{0.8215} \\
\num{11} & $[\num{10},\num{11})$ & \num{9} & \num{84} & $(\num{6},\num{13})$ & \num{0.9293} \\
\num{12} & $[\num{11},\num{12})$ & \num{13} & \num{149} & $(\num{13},\num{15})$ & \num{1.026} \\
\num{13} & $[\num{12},\num{13})$ & \num{12} & \num{150} & $(\num{13},\num{16})$ & \num{1.407} \\
\num{14} & $[\num{13},\num{14})$ & \num{17} & \num{150} & $(\num{13},\num{16})$ & \num{0.9411} \\
\num{15} & $[\num{14},\num{15})$ & \num{3} & \num{150} & $(\num{13},\num{16})$ & \num{0.9011} \\
\num{16} & $[\num{15},\num{16})$ & \num{4} & \num{84} & $(\num{6},\num{13})$ & \num{0.9648} \\
\num{17} & $[\num{16},\num{17})$ & \num{6} & \num{84} & $(\num{6},\num{13})$ & \num{1.037} \\
\num{18} & $[\num{17},\num{18})$ & \num{11} & \num{149} & $(\num{13},\num{15})$ & \num{1.426} \\
\num{19} & $[\num{18},\num{19})$ & \num{20} & \num{150} & $(\num{13},\num{16})$ & \num{1.03} \\
\num{20} & $[\num{19},\num{20})$ & \num{13} & \num{150} & $(\num{13},\num{16})$ & \num{0.9634} \\
\num{21} & $[\num{20},\num{21})$ & \num{11} & \num{150} & $(\num{13},\num{16})$ & \num{1.4} \\
\num{22} & $[\num{21},\num{22})$ & \num{19} & \num{150} & $(\num{13},\num{16})$ & \num{0.9872} \\
\num{23} & $[\num{22},\num{23})$ & \num{16} & \num{150} & $(\num{13},\num{16})$ & \num{0.9604} \\
\num{24} & $[\num{23},\num{24})$ & \num{14} & \num{150} & $(\num{13},\num{16})$ & \num{0.9541} \\
\num{25} & $[\num{24},\num{25})$ & \num{11} & \num{150} & $(\num{13},\num{16})$ & \num{1.4} \\
\num{26} & $[\num{25},\num{26})$ & \num{20} & \num{150} & $(\num{13},\num{16})$ & \num{1.03} \\
\num{27} & $[\num{26},\num{27})$ & \num{2} & \num{150} & $(\num{13},\num{16})$ & \num{0.8215} \\
\num{28} & $[\num{27},\num{28})$ & \num{7} & \num{84} & $(\num{6},\num{13})$ & \num{0.9636} \\
\num{29} & $[\num{28},\num{29})$ & \num{8} & \num{149} & $(\num{13},\num{15})$ & \num{0.89} \\
\num{30} & $[\num{29},\num{30})$ & \num{14} & \num{149} & $(\num{13},\num{15})$ & \num{0.9821} \\
\bottomrule
\end{tabular}
\end{minipage}\hfill
\begin{minipage}[t]{.49\textwidth}
\centering
\scriptsize
\setlength{\tabcolsep}{3pt}
\begin{tabular}{@{}llllll@{}}
\toprule
Stage & Time (\si{\second}) & Atk node $f$ & Active $\kappa$ & Active edge & $\mathrm{tr}\,W_c$ \\ \midrule
\num{31} & $[\num{30},\num{31})$ & \num{20} & \num{150} & $(\num{13},\num{16})$ & \num{1.03} \\
\num{32} & $[\num{31},\num{32})$ & \num{12} & \num{150} & $(\num{13},\num{16})$ & \num{1.407} \\
\num{33} & $[\num{32},\num{33})$ & \num{17} & \num{150} & $(\num{13},\num{16})$ & \num{0.9411} \\
\num{34} & $[\num{33},\num{34})$ & \num{1} & \num{150} & $(\num{13},\num{16})$ & \num{0.9753} \\
\num{35} & $[\num{34},\num{35})$ & \num{5} & \num{84} & $(\num{6},\num{13})$ & \num{1.253} \\
\num{36} & $[\num{35},\num{36})$ & \num{8} & \num{84} & $(\num{6},\num{13})$ & \num{0.8043} \\
\num{37} & $[\num{36},\num{37})$ & \num{15} & \num{149} & $(\num{13},\num{15})$ & \num{1.074} \\
\num{38} & $[\num{37},\num{38})$ & \num{18} & \num{150} & $(\num{13},\num{16})$ & \num{0.8456} \\
\num{39} & $[\num{38},\num{39})$ & \num{2} & \num{150} & $(\num{13},\num{16})$ & \num{0.8215} \\
\num{40} & $[\num{39},\num{40})$ & \num{8} & \num{84} & $(\num{6},\num{13})$ & \num{0.8043} \\
\num{41} & $[\num{40},\num{41})$ & \num{13} & \num{149} & $(\num{13},\num{15})$ & \num{1.026} \\
\num{42} & $[\num{41},\num{42})$ & \num{7} & \num{150} & $(\num{13},\num{16})$ & \num{1.272} \\
\num{43} & $[\num{42},\num{43})$ & \num{11} & \num{149} & $(\num{13},\num{15})$ & \num{1.426} \\
\num{44} & $[\num{43},\num{44})$ & \num{16} & \num{150} & $(\num{13},\num{16})$ & \num{0.9604} \\
\num{45} & $[\num{44},\num{45})$ & \num{20} & \num{150} & $(\num{13},\num{16})$ & \num{1.03} \\
\num{46} & $[\num{45},\num{46})$ & \num{3} & \num{150} & $(\num{13},\num{16})$ & \num{0.9011} \\
\num{47} & $[\num{46},\num{47})$ & \num{8} & \num{84} & $(\num{6},\num{13})$ & \num{0.8043} \\
\num{48} & $[\num{47},\num{48})$ & \num{4} & \num{149} & $(\num{13},\num{15})$ & \num{0.9346} \\
\num{49} & $[\num{48},\num{49})$ & \num{6} & \num{84} & $(\num{6},\num{13})$ & \num{1.037} \\
\num{50} & $[\num{49},\num{50})$ & \num{11} & \num{149} & $(\num{13},\num{15})$ & \num{1.426} \\
\num{51} & $[\num{50},\num{51})$ & \num{17} & \num{150} & $(\num{13},\num{16})$ & \num{0.9411} \\
\num{52} & $[\num{51},\num{52})$ & \num{11} & \num{150} & $(\num{13},\num{16})$ & \num{1.4} \\
\num{53} & $[\num{52},\num{53})$ & \num{16} & \num{150} & $(\num{13},\num{16})$ & \num{0.9604} \\
\num{54} & $[\num{53},\num{54})$ & \num{14} & \num{150} & $(\num{13},\num{16})$ & \num{0.9541} \\
\num{55} & $[\num{54},\num{55})$ & \num{17} & \num{150} & $(\num{13},\num{16})$ & \num{0.9411} \\
\num{56} & $[\num{55},\num{56})$ & \num{2} & \num{150} & $(\num{13},\num{16})$ & \num{0.8215} \\
\num{57} & $[\num{56},\num{57})$ & \num{1} & \num{84} & $(\num{6},\num{13})$ & \num{0.9754} \\
\num{58} & $[\num{57},\num{58})$ & \num{8} & \num{84} & $(\num{6},\num{13})$ & \num{0.8043} \\
\num{59} & $[\num{58},\num{59})$ & \num{15} & \num{149} & $(\num{13},\num{15})$ & \num{1.074} \\
\num{60} & $[\num{59},\num{60})$ & \num{6} & \num{150} & $(\num{13},\num{16})$ & \num{1.71} \\
\bottomrule
\end{tabular}
\end{minipage}
\end{table*}

%% file: adp_recovery_proofs.tex
\subsection{Proof of Lemma~\ref{lem:bellman}}
For any two tables $Q$ and $\widetilde Q$,
\begin{align*}
&|(\mathcal TQ)(f,a)-(\mathcal T\widetilde Q)(f,a)|\\
&\quad\le\gamma\sum_{f'}p_a(f,f')
\left|\min_bQ(f',b)-\min_b\widetilde Q(f',b)\right|\\
&\quad\le\gamma\|Q-\widetilde Q\|_\infty.
\end{align*}
Taking the maximum over $(f,a)$ proves contraction.
The finite table space is complete, so the contraction mapping theorem gives a unique fixed point.
Setting $V(f)=\min_aQ^\star(f,a)$ in the fixed-point equation gives \eqref{eq:bellman_exact}, whose unique solution is $V^\star$. The minimizing actions are therefore optimal.
Taking $\widetilde Q=Q^\star$ gives \eqref{eq:adp_exact_contraction}.
This is the discounted Bellman argument of \cite{Bertsekas2012}, specialized to the transition cost $J_c(f',a)$.

For the regret bound, choose $a^\star\in\arg\min_aQ^\star(f,a)$ and $\hat a\in\arg\min_a\hat Q(f,a)$.
Because $\hat Q(f,\hat a)\le\hat Q(f,a^\star)$,
\begin{align*}
Q^\star(f,\hat a)-Q^\star(f,a^\star)
&\le Q^\star(f,\hat a)-\hat Q(f,\hat a)\\
&\quad+\hat Q(f,a^\star)-Q^\star(f,a^\star)\\
&\le2\|\hat Q-Q^\star\|_\infty.
\end{align*}
This proves \eqref{eq:adp_regret_bound}.\hfill$\square$

\subsection{Proof of Proposition~\ref{prop:adp_recovery}}
Replacing $J_c$ by $\bar J_c$ leaves the preceding contraction argument unchanged.
The corresponding operator $\bar{\mathcal T}$ therefore has a unique fixed point $\bar Q^\star$.
Conditional on $\mathcal H_k$, the table $\hat Q_k$ and the pair $(f_k,a_k)$ are fixed.
Taking expectations first over the cost given the realized next node and then over that node gives
\begin{align*}
&\mathbb E\!\left[c_k+\gamma\min_b\hat Q_k(f_{k+1},b)\mid\mathcal H_k\right]\\
&\quad=\sum_{f'}p_{a_k}(f_k,f')
\left[\bar J_c(f',a_k)+\gamma\min_b\hat Q_k(f',b)\right]\\
&\quad=(\bar{\mathcal T}\hat Q_k)(f_k,a_k).
\end{align*}
Thus the update noise
\[
M_{k+1}=c_k+\gamma\min_b\hat Q_k(f_{k+1},b)
-(\bar{\mathcal T}\hat Q_k)(f_k,a_k)
\]
has zero conditional mean, and the sampled update is
\begin{align*}
\hat Q_{k+1}(f_k,a_k)
&=(1-\alpha_k)\hat Q_k(f_k,a_k)\\
&\quad+\alpha_k(\bar{\mathcal T}\hat Q_k)(f_k,a_k)
+\alpha_k M_{k+1}.
\end{align*}
The bounded conditional second moments of $c_k$ imply a noise bound of the form
\[
\mathbb E[M_{k+1}^2\mid\mathcal H_k]
\le C_0+C_1\|\hat Q_k\|_\infty^2
\]
for finite constants $C_0,C_1$.
The finite state and action sets, contraction, nonanticipating sampling, and per pair learning rate conditions are the hypotheses of the asynchronous stochastic approximation convergence theorem for discounted tabular Q-learning \cite[Theorem~3 and Section~7]{Tsitsiklis1994Qlearning}.
It follows that \eqref{eq:adp_convergence} holds.

For any table $Q$,
\[
\|\bar{\mathcal T}Q-\mathcal TQ\|_\infty\le\delta.
\]
At the two fixed points this gives
\begin{align*}
\|\bar Q^\star-Q^\star\|_\infty
&\le\|\bar{\mathcal T}\bar Q^\star-\bar{\mathcal T}Q^\star\|_\infty
+\|\bar{\mathcal T}Q^\star-\mathcal TQ^\star\|_\infty\\
&\le\gamma\|\bar Q^\star-Q^\star\|_\infty+\delta.
\end{align*}
Rearranging yields \eqref{eq:adp_bias_bound}.
For $\delta=0$, $\bar Q^\star=Q^\star$.
To verify recovery of optimal decisions, consider the finitely many strictly suboptimal state-action pairs.
If any exist, their smallest gap
\[
\Delta=\min_{(f,a):\,Q^\star(f,a)>V^\star(f)}
\left[Q^\star(f,a)-V^\star(f)\right]
\]
is positive.
Almost sure convergence eventually makes $\|\hat Q_k-Q^\star\|_\infty<\Delta/2$, so the regret bound rules out every strictly suboptimal greedy action.
If no such pairs exist, every action is already optimal.
With ties, the greedy action can vary within the optimal set.\hfill$\square$